# DIGITAL RESILIENCE FOR WHAT? CASE STUDY OF SOUTH KOREA

Kyung Ryul Park, Korea Advanced Institute of Science and Technology, park.kr@kaist.ac.kr

Sundeep Sahay, University of Oslo, sundeeps@ifi.uio.no

Jørn Braa, University of Oslo, jbraa@ifi.uio.no

Pamod Amarakoon, University of Colombo, pamod@pgim.cmb.ac.lk

**Abstract:** Resilience has become an emerging topic in various fields of academic research. In spite of its widespread use, there remains conceptual confusion over what resilience means particularly in multi-disciplinary studies including the field of ICT and Development. With the potential of digital technology, research is needed to critically question what key socio-institutional values related to resilience are being strengthened, for what and for whom through the different conceptualizations of resilience. In this study, we conduct an interpretive case study on South Korea's response to the pandemic and construct a chronological narrative to identify key aspects of digital resilience. We identify agility, diversity, and plurality - enabled by active roles of various stakeholders, including citizens, research communities, and private sector - as keys to digital resilience to the pandemic. Findings from the case of South Korea provide implications to ICT4D research while discussing how developing countries, where a national single window platform is typically implemented with greater level of homogeneity, achieve digital resilience with inclusive innovation with plurality of diverse platforms.

**Keywords:** COVID-19, pandemic, resilience, digital technology, agility, citizen engagement, public private partnership, R&D, governance, South Korea

## 1. INTRODUCTION

South Korea's response to the COVID-19 pandemic has been considered by international authorities and media as one of the world's most effective (World Health Organization, 2020). The Korean government comprehensively summarized how the country has taken various digital interventions to "flatten the curve" of COVID-19 (Ministry of Economy and Finance, 2020), including for the testing, tracing and treating of COVID-19 cases. The effectiveness of these interventions allowed the government to avoid a lockdown, like most countries did (D. Lee, Heo, & Seo, 2020). Conventional contact tracing methods involved exchange of documents with the National Police Agency with GPS locations acquired from telecommunication companies (MSIT, 2020). However, advanced digital technologies supported by new data governance models were harnessed to support the public-health response integrating surveillance, contact tracing, and information sharing with the public. This led to the government's adaptive response and information sharing during the pandemic.

Lessons from the South Korea case provides important research and policy implications for digital resilience, including for building the following competencies: innovative and agile development of digital applications, efficient data governance, citizens' active engagement and public-private partnership. We analyse these competencies from the perspective of digital resilience, drawing from the experience of the government's information systems response to the pandemic and citizen's participation and uptake of these responses. The rest of the paper is organized as follows. Section 2 identifies related conceptual issues. Section 3 presents research design, methods, and case description of South Korea information systems response to the pandemic. Section 4 offers a chronological analysis of the case, followed by discussion in Section 5.

## 2. CONCEPTUAL FRAMEWORK

Resilience has become a popular concept used by enterprise, government and international organizations, and in research in many disciplines such as Ecology, Management, Disaster Studies, Development Studies and Information Systems. IS research has mostly studied resilience in the context of organizations, infrastructures and risk management, with a focus on identifying attributes of resilience such as agility, flexibility, adaptability, governance, and decreased vulnerability (Erol, Sauser, & Mansouri, 2010; Heeks & Ospina, 2019). In spite of its widespread discussion, there remains conceptual confusion over what resilience means particularly in the context of ICT for Development (ICT4D). Heeks and Ospina ask the question: ‘resilience of what? whether it is about the information systems as an object or the wider socio-institutional system as a target?’ Related relevant questions include ‘resilience for whom?’ and ‘resilience at which institutional level?’

The meaning of resilience varies at different institutional levels: such as i) mental and emotional resilience, e.g., trust, cognitive flexibility and diversity, at individual and community levels; ii) relational resilience, e.g., multidisciplinary teams, interaction and collaboration, at organizational level; and finally, iii) societal resilience, e.g., curative action/vaccine, across border solidarity and collaboration, at national and global levels. These three levels emphasize the inter-connections of aspects of trust, transparency, collaboration, networking, solidarity and institutional systems (Rawat, Boe, & Piotrowski, 2020) and also about regional innovation and networking (Bristow & Healy, 2014; Bristow & Healy, 2018).

In regional economy, the notion of complex adaptive systems provide insights into resilience and how the system rearranges its internal structure spontaneously in response to an external shock or to some internal or ‘self-organized criticality’ (Lansing, 2003). Building on this notion, Braa et. al. (2007) develop the lens of duality between ‘bounce back’ and ‘bounce forward’, and relate them to notions of adaptation and adaptability through their analysis of health data standardization and the role of attractors. While adaptation reflects a path-dependent process maintaining existing paths or primary functions of a system, adaptability refers to an adaptive ability, in pursuit of new path creation and structural change (Hu & Hassink, 2015).

Wilson (2012) provides a framework for community resilience understood as the intersection between economic, social and environmental components. The resilience vs vulnerability dichotomy can be expressed as a spectrum, where the extreme ends are easily conceptualized. Resilience is thus understood as an ideal end goal, whilst acknowledging that no community is totally resilient or vulnerable but will in most cases contain elements of both, raising the need to study it not as an outcome but as a process (Wilson, 2012).

Bruneau et al. (2003) conceptualize resilience along four interrelated dimensions of technical, organizational, social, and economic, with the following attributes: i) robustness, e.g., ability to withstand a given level of stress without suffering degradation or loss of function; ii) redundancy, e.g., the extent to which elements, systems, or other units of analysis exist that are substitutable; iii) resourcefulness, e.g., capacity and ability to mobilize resources; and, iv) rapidity, the speed of response. While Heeks and Ospina (Heeks & Ospina, 2019) emphasize additional attributes of self-organization, learning, equality, flexibility and scale, Erol et al (2010) adds increased adaptability and decreased vulnerability, representing process mechanisms in achieving resilience.

This conceptual review has shown that resilience means different notions to different research fields and there is the need to establish conceptual clarity and framework for transdisciplinary research such as ICT4D. In order to identify key attributes of resilience, we focus on the following key aspects: innovative applications of digital technologies, agility, efficient data governance, citizens’ active engagement, public-private partnership (PPP), resourcefulness, and research and development (R&D) capacity.

# 3. METHODOLOGY

## 3.1. Data Collection

In this study, we develop a single embedded interpretive case study to understand digital response to the pandemic in the context of South Korea. Data collection involved conducting in-depth interviews with two government officers, one developer of a citizen-oriented application, and one health policy expert within Korean academia. These interviews were semi-structured and focused on understanding: i) the cases and details of ICT applications used in the context of the COVID-19 response; ii) government's motivations for using the different digital solutions; iii) the challenges experienced in deploying the different digital solutions; and, iv) self-evaluation of the effectiveness of the ICT tools used. Their responses helped the reconstruction of the case narrative from an information systems perspective. With the exception of one in-person interview, the others took place online and over the telephone, and lasted between 45 minutes and one and a half hours. Two interviews (developer and health expert) were audio-recorded and transcribed in Korean, while permissions for recording the interviews with the two government officers were not granted. A variety of secondary publications were studied, including: i) official government reports; ii) open data portals and other citizen generated applications; iii) publications from conferences and journals; and, iv) media articles. A summary of some of the publications is provided in the table below.

| | |
|---|---|
| Government's official publications | · Government of the Republic of Korea. (2020). Tackling COVID-19 Health, Quarantine and Economic Measures: Korean Experience (Sejong: March 31, 2020); Government of the Republic of Korea. (2020). Flattening the curve on COVID-19: How Korea responded to a pandemic using ICT (Sejong: April 15, 2020). These two documents are the very first government official reports which different ministries jointly contributed including Ministry of Economy and Finance (MOEF), Ministry of Science and ICT (MSIT), Ministry of the Interior and Safety (MOIS), Ministry of Health and Welfare (MOHW), Ministry of Land, Infrastructure and Transport (MOLIT) and Korean Intellectual Property Office)<br>· Central Disaster and Safety Countermeasure Headquarters. (2020). Korea's Response to COVID-19 and Future Direction.<br>· Ministry of Science and ICT. (2020). How We Fought COIVD-19: A perspective from Science and ICT. |
| Open data portals | · Ministry of Health and Welfare (MOHW). Coronavirus Disease 2019. http://ncov.mohw.go.kr/en<br>· Korea Center for Disease Control (KCDC). Covid-19 Updates. http://www.kdca.go.kr |
| Publicly available conferences presentations and discussions | · Ministry of Land, Infrastructure and Transport (MOLIT) and KCDC (2020). Corona19 smart management system online briefing (April 10, 2020). https://www.youtube.com/watch?v=C9o_HGN6v8E&feature=emb_logo |
| Citizens' generated applications | · Corona Map: https://coronamap.site<br>· Corona Live: https://www.coronalive.co.kr |

**Table 1. Summary of publications referred for case study in South Korea**

The combinations of the primary and secondary data were consolidated and themes related to resilience were identified to develop more generalized insights for discussion.

### 3.2. Case Description

In late December 2019, there were reports of pneumonic patients arising in Wuhan, China. As a result, KCDC (Korea Center for Disease Control) enhanced the quarantine and screening measures for travelers entering from Wuhan at points of entry, with the cooperation of the World Health Organization (WHO) and the Chinese health authorities. The first patient confirmed as a case of COVID-19 in Korea was announced on January 20, 2020. The carrier arrived in Incheon International airport from Wuhan, China, and tested positive after being flagged for having high temperature at the entry screening using an infrared sensor. Korea was successfully able to control and mitigate the spread of the epidemic without the need for a lockdown or restrictions on movement. The Korean national crisis and emergency management system consists of four alert levels, on which the government raised the infectious disease alert levels to Blue (Level 1) on January 3$^{rd}$; to Yellow (Level 2) on January 20$^{th}$, to Orange (Level 3) on January 28$^{th}$, to Red (Level 4) on February 23$^{rd}$ (MOHW, 2020), as summarized in Table 2.

| Month | Agile Adaptive Policy Decision | Information Systems Responses |
|---|---|---|
| **2020**<br><br>**Jan** | · Quarantine and Screening measures are enhanced for inbounds entering from Wuhan – Infectious Disease alert category Blue (Level 1) (January 3)<br>· Cooperation of KCDC, Ministry of Interior and Safety (MOIS), and the Ministry of Justice (MOJ) (January 8)<br>· The Korean government changes its Infectious Disease alert category Blue (Level 1) to Yellow (Level 2) (January 20)<br>· First positive COVD-19 patient in Korea (January 20)<br>· KCDC extended the definition/ case of the novel coronavirus infected cases and strengthened monitoring (January 26)<br>· Infectious Disease alert category from Yellow (Level 2) to Orange (Level 3) (January 28) | · Immigration Screening (January 8)<br>· KCDC introduce '1339 Call Center' (January 29)<br>· Pharmacies are given permission to check patients' travel histories through status checking system (January 31) |
| **Feb** | · MOHW denies entry of foreigners from Hubei Region - Korean citizens returning from Hubei region or those who have been in contact with positive tested patients must self-isolate for 14 days from the date of entry (February 2)<br>· MOHW release guideline on operation for group facilities<br>· Temporarily close if any occupant is tested positive for COVID-19 (February 3)<br>· Public and Private health facilities designated for additional testing areas (February 7)<br>· KCDC announces quarantine screening and measures extended to travellers from China, Hong Kong, and Macao (February 12)<br>· MOHW and KCDC dispatch special task forces to Daegu region implement disease control measures (February 19) | · PCR test kits available (Kogene Biotech) at Incheon International airport (February 4)<br>· Self-Diagnostic app developed (February 12)<br>· Introduction for contact mapping application and the initiation of using mobile tower for GPS location for contact mapping. Contact Mapping Application & Mobile Tower Location (February 18) |
| | · Infectious Disease alert category Orange (Level 3) to Red (Level 4)<br>· Delayed new school year by 1 week (February 23)<br>· Drive-thru testing checkpoint operated by local government (February 26)<br>· KCDC advises social distancing and maintain hygiene measures (February 28) | · Telemedicine initiation (February 23)<br>· Social media and other platforms used for election campaigning (February 28) |
| **Mar** | · Delayed new school year by 3 weeks (March 2)<br>· Korean government supply mask on a 5-day rotation allowing two masks purchase per person (March 5)<br>· Mask Map service started (March 11)<br>· Entry restrictions for inbound from 76 countries (March 16)<br>· K-Walk thru testing station set up outdoors at Incheon International airport (March 26) | · GPS-based app for self-quarantine measures (March 7)<br>· Official start of Epidemiological Investigation Supporting System (EISS) (March 26) |
| **Apr** | · MOHW announce a fine up to 10 million Korean won (~$8,000) or imprisonment for those who not comply to 14 days self-quarantine measures (April 5)<br>· Introduction of Self-quarantine Safety Band (April 27) | · Introduction of so-called '*untact'*[1] services which range from Artificial Intelligence (AI) food/beverage ordering machine, online shopping to ordering food remotely and citizens' participation (April 9) |
| **May** | · MOLIT require all passengers to wear masks on flight and public transportation (May 25) | |
| **Jun-** | | · QR code-based registration at restaurants and entertainment facilities (June 10) |

**Table 2. Key milestones**

With massive applications of digital technologies, Korea initiated quarantine measures and most significantly the testing for COVID-19 in the initial stages. People with history of contacts with proven cases were strictly quarantined, with digital monitoring of the quarantine. The effective response against a novel infectious disease required expertise knowledge in order to implement the required strict measures (Shaw et al., 2020). The Korean government had assembled the Central Disaster and Safety Countermeasure Headquarters which is headed by the Prime Minister to reinforce the government's agile responses. In fact, specific roles and responsibilities were allocated such that the KCDC Headquarters served as the command center of the prevention and control effects, the MOHW, head of the Central Disaster Management Headquarters, assisted the KCDC. Meanwhile, the MOIS, head of Pan-government Countermeasures Support Headquarters, liaised and coordinated assistance between the central and local governments (MOIS, 2020). Nonetheless, the administered system and health policies were not possible without the help of digital technologies. Heo et al. (2020) argue that ICTs allowed the Korean government to endorse containment and mitigation strategies simultaneously such as warning alerts, epidemiological investigation, masks-wearing, quarantine contact, and social distancing. In late January, the KCDC introduced the '1339 Call Center' which provides counselling and guidelines for suspected persons symptomatic of COVID-19 and also to check the closest screening station (MOHW, 2020). Furthermore, the Korean government was able to effectively equip itself with robust health and quarantine measures using the 3Ts modules, i.e., test, trace, and treat (Lee et al. 2020). The following section presents a chronological analysis of the case.

[1] Korean English word similar in meaning to 'non-contact'; a newly coined word of "contact" with a negative prefix "un", referring to non-face-to-face social and economic activities facilitated by advanced information communications technology (ICT). See the explanation on Section 5.3.

# 4. CASE ANALYSIS

In order to analyse the case, this section presents a narrative chronological account of the response to the COVID-19 and the roles of digital technologies. The case is embedded in the learning-based path dependence of government's previous pandemic response. We turn to the government's response, setting out briefly some of the historical environment that provides an opportunity to build adaptation and adaptability in Korean government.

***Lessons from prior pandemics:*** Just as with any other crisis, pandemic response requires centralised leadership and a clear role in the chain of command. Korea became sensitive and adaptive to the dangers of similar viruses through its experience with prior pandemics of the Severe Acute Respiratory Syndrome (SARS) in 2003 and the Middle East Respiratory Syndrome (MERS) in 2015. The lack of emergency preparedness and response led to the rapid spread of MERS to patients, visitors, and healthcare workers at the hospitals, which also suffered from poor ventilation and lack of emergency training to healthcare workers (Ha, 2016). Consequently, the Korean government opted for non-transparency to avoid citizens' panic, which caused friction between the central and local governments especially in the capital city of Seoul, where the latter wanted more transparency (Moon, 2020). Kim et al. (2020) also identified the issue of poor administrative and data coordination between the MOHW and the subordinate organisation of MOHW, KCDC, which contributed to the system failure of national disease control and to the slow response. The Minister of Public Safety and Security (MPSS, which later became Ministry of Interior and Safety, MOIS), being another important line ministry, did not implement specific response to prevent deaths (Ha, 2016). The painful failure in eradicating the MERS outbreak prompted the government to act expediently to build agile responses and commands during the outbreak of COVID-19 in early 2020, and paved a way to more resilient responses from the government.

***Blue level alert (December 2019 – January 2020)***: In December 2019, the Korean government instructed the Blue Alert Level (the lowest level of alert in the national crisis management system) as the first case of novel Covid-19 was reported. The government assembled an emergency response team within a day that included the Central Disaster and Safety Countermeasure Headquarters headed by the Prime Minister, with the KCDC as the command centre, with the MOHW is close support. The MOIS liaised and coordinated assistance between the central and local governments (Ministry of Health & Welfare : News & Welfare Services, 2020). From the historical experiences of SARS and MERS, biotechnological companies were able to rapidly create reagents and kits for testing COVID-19. The real-time reverse transcription polymerase chain reaction (rRT- PCR) diagnostic sequencing method was shared by KCDC which partnered with the Korean Society for Laboratory Medicine (KSLM) and the Korean Association for External Quality Assessment Service (KAEQAS) in January 2020 (You, 2020).

***Yellow Level – Containment (January 2020)*****:** As soon as the first positive case was reported on January 20 at Incheon airport, the disease alert category was increased to Yellow (Level 2), and the focus shifted on ICT-based mass testing and the Central Discharge Countermeasures Headquarters (CDCHQs) was set up to initiate the 24-hour emergency response system. Quarantine and screening measures were set up for all individuals traveling to Korea within 14 days of visiting Wuhan, China, and were required to submit to the health questionnaire and report any fever or respiratory symptoms experienced. Symptoms of those entering Korea were monitored by using a self-quarantine safety app which registered quarantine location. The central and local governments established coordination structure to strengthen process to assess the situation, disclose information, provide transparency and built coordinated collective action.

***Orange Level – Contact Tracing (January 27-February 17)*****:** On January 28, to strengthen community response, the level was raised to Orange (Level 3) and all incoming people were subject to 14 days of self-quarantine. The government endorsed public-private partnerships with advanced ICT and biotech companies (Kogene Biotech, Seegene, Solgent, SD Biosensor, and Bioseworm) to ramp up testing by developing COVID-19 diagnostic kits which were made available at airports on February 4$^{th}$ (The Government of Republic of Korea, 2020). As the number of cases started to increase, the central and local authorities proposed the deployment of a contact tracing application for all cases entering the country. The immediate response team would investigate on potential flareups at diverse locations to investigate traces of confirmed cases. Those confirmed was required to install the "Self-Diagnosis App" and "Self-Quarantine Safety Protection App" allowing users to monitor their health conditions and access readily available information on follow-up actions such as physical check-ups, and access to healthcare services using the 1399 line.

***Red Level – Epidemiological Investigation (February 18-)*****:** On February 18$^{th}$, patient 31 attended service at the Shincheonji Church, and as a result contributed to the cluster of COVID-19 cases in the days to follow, putting the country on Red Level Alert. The government implemented a 5-day rotation system allowing two masks to be purchased per person on designated days based on their birth year. In the efforts to minimize the spread of COVID-19 between healthcare workers and citizens, the government initiated new types of screening stations which are called Drive-thru and K-Walk-thru. Further, leveraging on the advanced ICTs, the MOLIT and the KCDC selected the Smart City data platform and jointly developed the Epidemic Investigation Support System (EISS) to provide an infographic mapping of those tested positive (who) and their routes of the infection (where), to estimate risks which were visualized by hot-spot analysis (Park *et al.*, 2020). AI based medical imaging (including x-ray and CT scan) analysis devices were created to speedily detect major lung abnormalities (Ministry of Science & ICT - Republic of Korea, 2020).

***Global Pandemic Level – Citizen participation (March-)*****:** Since March, the government has been making efforts in curbing the contagion through the Cellular Broadcasting Service (CBS) to inform the public of the movement routes taken by confirmed COVID-19 patients. The government used mass media and various channels to promote the level of awareness and encourage citizens' participation and promote '*untact*' behaviors such as ordering food remotely using delivery applications, and making online schooling available. Also on July 12$^{th}$, public mask system was abolished and shifted to market supply system. There is also an institutional change in health governance. On September 12$^{th}$, it was announced that the KCDC has underwent an institutional reform, providing itself the independent power over budget and personnel from MOHW.

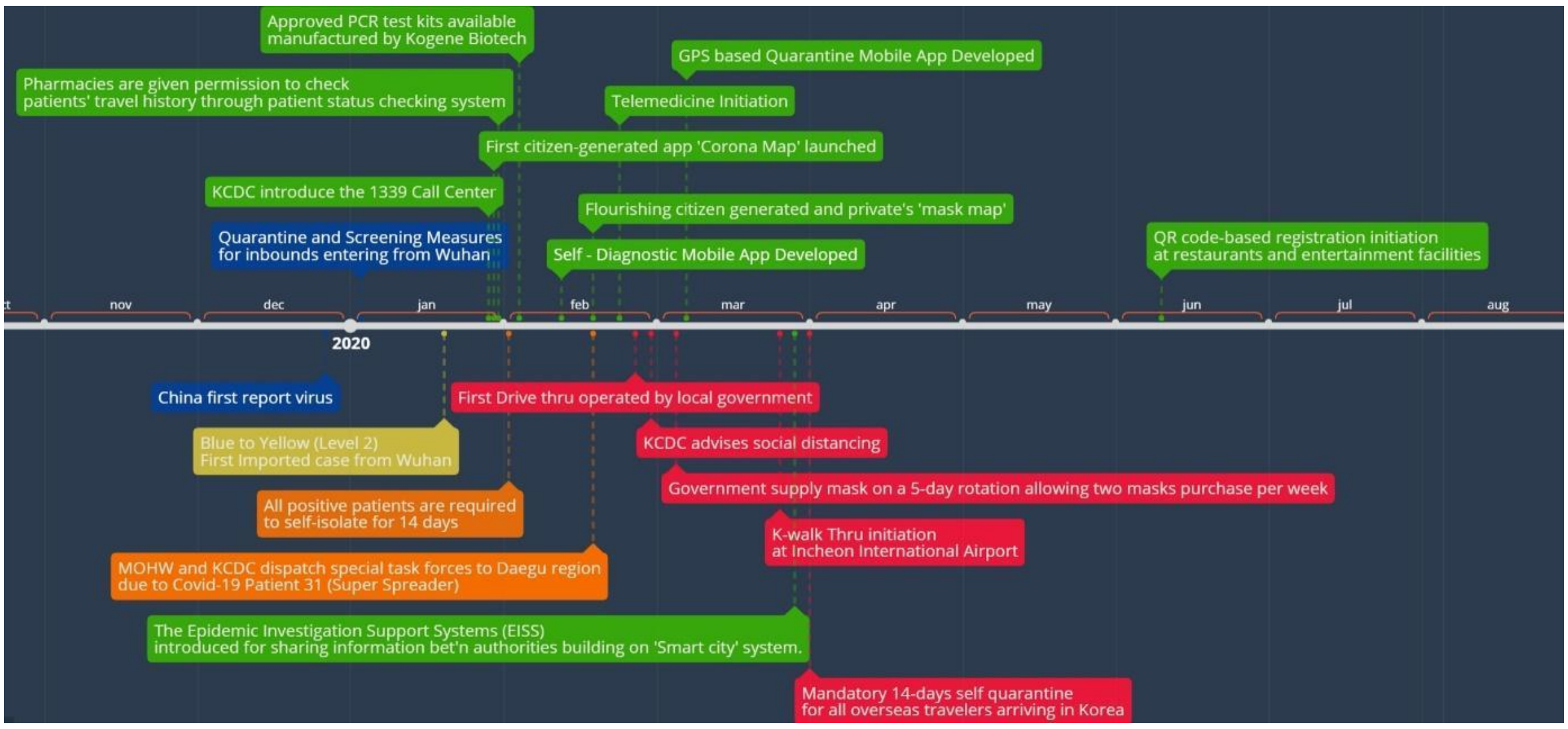

**Figure 1. Timeline of COVID-19 Surveillance Response in Korea**

Korea was one of the first countries to experience a COVID-19 case on 20th January 2020, after China (Shaw, Kim, & Hua, 2020). However, Korea was relatively successful to control the epidemic without the need for a lockdown or restrictions on movement (Lee & Lee, 2020). Although new confirmed cases increased up to 1,237 confirmed case per day during the Christmas season in 2020, overall daily new confirmed cases are around 600 in the last two months (April and May 2021). On February 26th, 2021, South Korea began vaccinating people aged 65 and older.

## 5. DISCUSSION

This section presents key enablers to digital resilience in the context of South Korea. We identify agility, diversity and plurality enabled by various digital actors' engagement such as citizen participation, private public partnership (PPP) and R&D governance. This section consists of the following discussion: role of digital technologies and platforms, agility, citizen engagement, PPP and R&D capacity as below.

### 5.1. Utilization of Digital Technologies

The extensive use of digital innovations based on advanced ICTs was fundamental in shaping the response, implemented within a strong legal framework to maintain standards for privacy. ICTs were used to share and improve the inventory of mask supplies and for citizens to communicate with government authorities using mobile apps. MOLIT developed the Smart Management System (SMS) in collaboration with the KCDC and the MSIT (Ministry of Science & ICT - Republic of Korea, 2020). The SMS collects big data, such as mobile phone location, hot spot usage, CCTV recordings, and credit card usage within 10 minutes (Ministry of Health & Welfare : News & Welfare Services, 2020). Artificial Intelligence (AI) played a significant role in supporting the diagnosis and screening of patients, and develop thorough analysis of different situations. AI based medical imaging (including x-ray and CT scan) analysis devices were created to detect major lung abnormalities, and assist doctors in making quick diagnosis (Ministry of Science & ICT - Republic of Korea, 2020).

The EISS performed inter-institutional communication across the KCDC, police departments, and mobile network companies analysing real time data using GPS and mobile data and CCTV footage, which allowed tracking of patient routes ("KCDC," 2020; Y. J. Park et al., 2020). The EISS is loaded on the cloud system TOAST G, providing access information of confirmed cases and the AI identifies any mismatched information such as location and time period of distant base stations (Y.

J. Park et al., 2020). Whereas the earlier system took 72 hours for contact tracing, the EISS could exchange information across 28 institutions in less than an hour (Y. J. Park et al., 2020). Moreover, the accuracy of epidemiological investigations was enhanced through data of patient's travel and medical history, provided by the Korean Immigration Service and the Health Insurance Review and Assessment Service (Ministry of Science & ICT - Republic of Korea, 2020). To safeguard patient privacy, the EISS was equipped from hacking and log-in access protocol. To protect personal information, the EISS limited access of data to epidemiology investigators to identity routes of only positively tested cases who were promptly quarantined (Y. J. Park et al., 2020).

Existing hospital information systems were enhanced with dashboards, e-prescriptions and cloud-based image sharing. New mobile apps enabled efficient communications between healthcare workers and patients, and monitoring devices facilitated in effective patient management (Bae et al., 2020). Quarantined patients were monitored by using a self-quarantine safety app, which registered location and alerted authorities when patients went out of their location. Self-diagnostic apps were made mandatory to those entering the country (Ministry of Health & Welfare : News & Welfare Services, 2020), and patients were identified as moderate, severe, and extremely severe, and appropriately admitted to national designated treatment facilities. Asymptomatic but positively tested patients were isolated and placed in government-sponsored facilities called Living and Treatment Support Centers (LTSCs), which helped prevent shortages of hospital beds. Doctors engaged in telemedicine services to avoid group contagion and contain the spread of infection (Heo, Lee, Seo, & Choi, 2020).

## 5.2. Agility

Learning from the MERS outbreak, the government had significantly ramped up its testing capacity drawing upon all available resources and supported by advanced ICTs to geo-locate positive cases (Moon, 2020; Y. J. Park et al., 2020). The government enacted a policy on emergency use authorization which enabled the use of pre-approved diagnostic kits in conditions of emergency. The agility was built by the use of multiple apps for different purposes, each built through multiple actors and partnerships. Different forms of learnings contributed to providing the basis for agility in both target and object system responses, which were socio-technical in nature (Amarakoon, Braa, Sahay, & Hewapathirana, R. Siribaddana, 2020). Our study reinforces the contribution of socio-technical agility in innovation and adaptability during the period when there is the need for rapid institutional change. From this perspective, it is notable that each local government established a Local Disaster and Safety Management Headquarters to secure the capacity of healthcare services and beds (Ministry of Foreign Affairs- Republic of Korea, 2020). The government established drive-thru and walk-thru screening stations by late February which shortened testing time and scaled testing capacity nationally, without putting health workers at risk (S. M. Lee & Lee, 2020; Ministry of Health & Welfare, 2020; Moon, 2020). Tests lasted 10 mins on average, per person, whereas previous tests and registration lasted 30 mins (Ministry of Health and Welfare, 2020). Korea was capable of conducting over 23,000 diagnostic tests free of charge per day. Patients were identified as moderate, severe, and extremely severe, and admitted to national designated treatment facilities. Asymptomatic but positive cases were isolated and placed in government-sponsored facilities called LTSCs, where they received two check-ups a day and were immediately transferred to a hospital if symptoms are aggravated (Ministry of Foreign Affairs- Republic of Korea, 2020). The Information Management System provided status of patients and bed availability (Heo et al., 2020), and if countermeasures required excess local capacity, the central government supported necessary resources including supplies, beds and personnel (Ministry of Foreign Affairs- Republic of Korea, 2020). In-patient hospital care and testing costs was covered by the National Health Insurance (NHI) or government funding.

### 5.3. Citizens' Engagement and Citizen-generated Apps

Leveraging on the existing digital infrastructure, the government was able to develop mobile applications which enabled '*untact*' services and behaviour of citizens. "*Untact*" a combination of the prefix 'un' and the word 'contact,' describes the health-protective behaviour stemming from individuals in the perception of risk of infection (S. Y. Bae & Chang, 2020). The authors argue that the Health Belief Model (HBM) framework explains the individual's health protective actions during COVID-19 pandemic. In fact, HBM have been successfully applied such as in vaccination in health-promoting behaviours. The fear of the unknown novel Corona virus, especially because patients are asymptomatic and the infection transmission rate are so rapid, has promoted *untact* behaviour in daily lives among Korean citizens, while minimizing the possible risk of infection. For instance, online purchasing and payment, working from home, and AI kiosks at restaurants have become more and more common (Lee & Lee, 2020). The Korean authorities also created low-contact testing such as drive through test stations, which minimized exposure to healthcare workers.

There were also citizen's alternative digital responses to data sharing. A couple of notable Geographic Information System (GIS)-based applications and websites aimed to help people access and share information. Among them, 'Corona Map' and 'Mask Map' attracted many users as supplementary sources for information. After the first outbreak in Korea, the KCDC published the trajectory history of patients on its website. After 10 days of the first case, one college student developed a 'Corona Map' which shows the travel routes of confirmed cases on website map service that is accessible via mobile or web. It is based on open source called 'Open Street Map' (Ha, 2020a). The information on the map can be also found on the website provided by KCDC, but it is hard to be found and only written in texts. While the Corona Map has its advantage in intuitively displaying the information by visualization, its increasing use made the developer worry about the cost. However, on February 3$^{rd}$, Amazon Web Service (AWS) contacted him for supporting all the costs rising from running the server. The following day, Naver gave free upgrade for Naver Map Application Programming Interface (API), and on February 6$^{th}$, Kakao provided Kakao Destination Search API for free.

Korea was not the first country to collaborate with citizen developers to provide mask availability information. However, the citizen generated Mask Map service is worth discussing. There was also a close partnership with IT service providers. When pharmacies inform Health Insurance Review & Assessment Service (HIRA) of the status of mask sales, the sales data is provided to National Information Society Agency (NIA) who reproduces the data in the form of Naver Map of Kakao Map API, so that it can be used in apps or the web (Ministry of Science and ICT Press Report, 2020). It is disclosed by cloud services to prevent delay of the data display and provide flexible services. The cloud alliance, comprised of Naver, KT, NHN, Koscom, offers free cloud service for two months. Other GIS-based applications were also actively developed in civic hacking communities in Korea. Understanding public participation and how this participation transforms knowledge, society and social orders is crucial for analysing the way people implement technoscience in solving social problems. Citizen participation can suggest implications for the data policy in public health crisis, such as voicing their opinions in the situation.

### 5.4. Public-Private Partnerships

PPP were effectively leveraged to support different interventions in an agile manner. While being important in the defence part of resilience, agility may be even more important in the bounce forward attack and innovative adaptability. A key intervention was the EISS, which enabled information sharing between the credit card company and mobile network companies. Park *et al.* (Y. J. Park et al., 2020) state that about 2,000 daily cases were tracked by mobile tracking applications, which reduced the time of epidemiological investigation to 10 mins, using information based on latitude and longitude, as well as of visit duration (Y. J. Park et al., 2020; You, 2020). The government collaborated with private agencies in the making of COVID-19 treatments and vaccines. On April 16th, 2020, The Coalition for Epidemic Preparedness Innovations (CEPI) granted funding of $6.9 million to INOVIO to work with The Korea National Institute of Health (KNIH), which functions under the KCDC, and the International Vaccine Institute (IVI) for the development and trial testing of INOVIO's phase I/II INO-4800, a nucleic-acid based vaccine (IVI, 2020). The hyper-network environment comprising the government, businesses, and citizen's digital contributed to the government 'flattening the curve' by following the 3T policy (test, trace, and treat). Within this framework, people with a history of contacts with proven cases were strictly quarantined, with digital monitoring of the quarantine. Braa et al. (2004) have emphasized the role of 'networks of action' in ICT4D research. While they don't directly address the issue of resilience, they emphasize the heterogeneity of networks as an important determinant of building robustness and sustainability. From this perspective, the network in the case of Korea was relatively more formalized and in-country, coordinated by the MOHW, and built around the framework of PPPs. This networking was primarily in-country and place-based, as contrasted to the more global nature of developing countries where many international donors play roles. However, the case shows that networks are crucial in enabling adaptability in crisis and changing environments.

### 5.5. Research and Development Capacity

Another key aspect of the context relevant for the Covid-19 response was the government's intensive funding for R&D in the health sector and in building effective PPP. Resilience requires building capacity at multiple institutional levels of the individuals, organizations, health system, and governance at national level. It also requires multi-faceted capacity relating to, for example, investments and science and technology policy that facilitated R&D for testing methods, leveraging on digital technologies, use of new technologies of AI and machine learning, emergency and disaster management, and various others. For the period of 2013-2017, the government invested KRW 1.14 trillion ($1,032 million) (Korea Pharmaceutical Bio-Association, 2020). MOHW, MSIT, the Ministry of Agriculture, Food and Rural Affairs, invested heavily on R&D with KRW 189 billion ($171 million) in total for vaccines, and a total of KRW 177 billion ($160 million) for diagnostic test kits (Korea Pharmaceutical Bio-Association, 2020). R&D was continuously funded, and efficient mechanisms were established to improve health emergency management, since MERS in 2015. For example, Kogene Biotech received a total of KRW 23.1 billion for the development of diagnostic technology using Multiple PCR, whilst SolGent and SD Biosensor got a total of KRW 20.5 billion and KRW 8.5 billion respectively for the development of diagnosis technology for Zika viruses based on PCR testing (Ministry of Science & ICT - Republic of Korea, 2020). The push for R&D investment is a crucial aspect of building successful diagnostic kits/drugs, which led to agile responses during the first impact of the COVID-19 outbreak. With this backdrop, the companies mentioned above managed to develop and supplied test kits to national and local governments by early March 2020, given the short time period (KMFDS, 2020). The development of these diagnostic reagents led to a significant reduction of time for diagnosis, enabling active response, enabled by the use of AI developed PCR reagents which supported accurate and rapid results from a single test (Ministry of Science & ICT - Republic of Korea, 2020). It is also notable that the KCDC used emergency procedures to fast-track the development and approval for the test kits during the early stage of the pandemic response.

## 6. CONCLUSION (280)

Although, the empirical findings derived from the single embedded case need to be used for general implications with caution, this study provides implications for digital innovation in developing countries. This study first broadly contribute to ICT4D literature and scholarly efforts to examine the role of information systems in developing countries (Park & Li, 2017; Walsham & Sahay, 2006). Our study may also be of interest to ICT practitioners and experts in health section, as it provides policy implications that highlight key attributes of digital resilience. In particular, in this study of South Korea, agility, plurality and diversity in digital pandemic responses were crucial as key contributors to digital resilience. Numerous digital actors were involved and co-evolved through public private partnership as well as citizens' coproduction to governance. In the case of South Korea, technological and economic strength combined with R&D and PPP were drivers of digital responses. On the other hand, in most developing countries, a parallel multitude in digital response to the pandemic are based on a national single window approach leading to the platform homogeneity and dilemma; should one go for one platform – or not. As Gawer (2014:1242) refer to this as the economies of scope in innovation explaining "the cost of jointly innovating on product A and B is lower than the cost of innovating on A independently of innovating B". A recent study in Sri Lanka reveals that agility, plurality and diversity in digital responses were seen as key factors in responding to the pandemic as well (Amarakoon, 2020). But in this case, they are achieved to a large extent through one platform. It is crucial to investigate how developing countries overcome socio-institutional, technical and managerial challenges and enable innovation through a platform of multiple applications, agility, plurality and diversity for achieving digital resilience. Future research also can be conducted to further clarify the intrinsic characteristics associated with resilience during a pandemic from both an institutionalist view and comparative point of view.

## REFERENCES AND CITATIONS

Amarakoon, P., Braa, J., Sahay, S., & Hewapathirana, R. Siribaddana, P. (2020). Building Agility in Health Information Systems to Respond to the COVID-19 Pandemic: The Sri Lankan Experience. In IFIP International Federation for Information Processing.

Bae, S. Y., & Chang, P. J. (2020). The effect of coronavirus disease-19 (COVID-19) risk perception on behavioural intention towards 'untact' tourism in South Korea during the first wave of the pandemic (March 2020). Current Issues in Tourism, 19.

Bae, Y. S., Kim, K. H., Choi, S. W., Ko, T., Wook, C. J. M. P., Cho, B., … Kang, E. (2020). Information Technology-based management of clinically healthy COVID-19 Patients: Lessons from a living and treatment support center operated by seoul national university hospital. Journal of Medical Internet Research, 22(6).

Braa, J., Monteiro, E., & Sahay, S. (2004). Networks of action: sustainable health information systems across developing countries. MIS Quarterly, 28(3), 337–362.

Bristow, G., & Healy, A. (2014). Regional Resilience: An Agency Perspective. Regional Studies, 48(5), 923–935.

Bristow, Gillian, & Healy, A. (2018). Innovation and regional economic resilience: an exploratory analysis. Annals of Regional Science, 60(2), 265–284.

Bruneau, M., Chang, S. E., Eguchi, R. T., Lee, G. C., O'Rourke, T. D., Reinhorn, A. M., … Von Winterfeldt, D. (2003, November). A Framework to Quantitatively Assess and Enhance the Seismic Resilience of Communities. Earthquake Spectra.

Erol, O., Sauser, B. J., & Mansouri, M. (2010). A framework for investigation into extended enterprise resilience. Enterprise Information Systems, 4(2), 111–136.

Gawer, A. (2014). Bridging differing perspectives on technological platforms: Toward an integrative framework. Research Policy, 43(7), 1239–1249.

Ha, K. M. (2016, March). A lesson learned from the MERS outbreak in South Korea in 2015. Journal of Hospital Infection. W.B. Saunders Ltd.

Heeks, R., & Ospina, A. V. (2019). Conceptualising the link between information systems and resilience: A developing country field study. Information Systems Journal, 29(1), 70–96.

Heo, K., Lee, D., Seo, Y., & Choi, H. (2020). Searching for digital technologies in containment and mitigation strategies: Experience from south korea covid-19. Annals of Global Health, 86(1), 1–10.

Hu, X., & Hassink, R. (2015). Overcoming the Dualism between Adaptation and Adaptability in Regional Economic Resilience. Papers in Evolutionary Economic Geography (PEEG).

Kim, Y., Oh, S. S., & Wang, C. (2020). From Uncoordinated Patchworks to a Coordinated System: MERS-CoV to COVID-19 in Korea. American Review of Public Administration, 50(6–7), 736–742.

KMFDS. (2020). News Briefing. Emergency approval for COVID-19 test kits.

Korea Pharmaceutical Bio-Association. (2020). KPBMA Policy Brief. Policy Report of the Korea Pharmaceutical Bio-Association, 20.

Lansing, S. (2003). Complex Adaptive Systems. Annual Review of Anthropology, 32, 183–204.

Lee, D., Heo, K., & Seo, Y. (2020). COVID-19 in South Korea: Lessons for developing countries. World Development, 135, 105057.

Lee, S. M., & Lee, D. H. (2020, March). "Untact": a new customer service strategy in the digital age. Service Business. Springer.

Ministry of Economy and Finance. (2020). Flattening the curve on COVID-19: How Korea responded to a pandemic using ICT. Sejong. Retrieved from http://www.korea.kr/common/download.do?fileId=190536078%0A&tblKey=GMN.

Ministry of Foreign Affairs- Republic of Korea. (2020). Korea's Response to COVID-19 and Future Direction View|Key Strategies.

Ministry of Health & Welfare : News & Welfare Services. (2020). Press Release-COVID-19 Response Meeting Presided Over by the Prime Minister.

Ministry of Science & ICT - Republic of Korea. (2020). How We Fought COVID-19: Perspective from Science & ICT.

Moon, M. J. (2020). Fighting <scp>COVID</scp> -19 with Agility, Transparency, and Participation: Wicked Policy Problems and New Governance Challenges. Public Administration Review, 80(4), 651–656.

Park, K. R., & Li, B. (2017). System failure for good reasons? Understanding aid information management systems (AIMS) with Indonesia as state actor in the changing field of aid. In International Conference on Social Implications of Computers in Developing Countries: IDT4D (Vol. 504, pp. 321–332). Yogyakarta, Indonesia: Springer.

Park, Y. J., Cho, S. Y., Lee, J., Lee, I., Park, W. H., Jeong, S., … Park, O. (2020). Development and utilization of a rapid and accurate epidemic investigation support system for covid-19. Osong Public Health and Research Perspectives, 11(3), 118–127.

Rawat, S., Boe, O., & Piotrowski, A. (2020). Military Psychology Response to Post Pandemic Reconstruction. Rawat Publications.

Shaw, R., Kim, Y., & Hua, J. (2020). Governance, technology and citizen behavior in pandemic: Lessons from COVID-19 in East Asia. Progress in Disaster Science.

The Government of Republic of Korea. (2020). Flattening the curve on COVID-19: How Korea responded to a pandemic using ICT.

Walsham, G., & Sahay, S. (2006). Research on information systems in developing countries: Current landscape and future prospects. Information Technology for Development, 12(1), 7–24.

Wilson, G. A. (2012). Community resilience and environmental transitions. Community Resilience and Environmental Transitions. Taylor and Francis.

World Health Organization. (2020). Sharing COVID-19 experiences: The Republic of Korea response. Retrieved from https://www.who.int/westernpacific/news/feature-stories/detail/sharing-covid-19-experiences-the-republic-of-korea-response

You, J. (2020). Lessons From South Korea's Covid-19 Policy Response. American Review of Public Administration, 50(6–7), 801–808.

## Appendix: List of Abbreviation

| | |
|---|---|
| 3T | Test, Trace, and Treat |
| AI | Artificial Intelligence |
| API | Application Programming Interface |
| CDCHQs | Central Discharge Countermeasures Headquarters |
| CEPI | Coalition for Epidemic Preparedness Innovations |
| COVID-19 | Coronavirus Disease 2019 |
| EISS | Epidemic Investigation Support System |
| GIS | Geographic Information System |
| GPS | Global Positioning System |
| HBM | Health Belief Model |
| HIRA | Health Insurance Review & Assessment Service |
| ICT | Information and Communications Technology |
| ICT4D | Information and Communications Technology for Development |
| IS | Information Systems |
| IVI | International Vaccine Institute IVI |
| KCDC | Korea Center for Disease Control |
| KNIH | Korea National Institute of Health KNIH |
| LTSCs | Living and Treatment Support Centers |
| MOEF | Ministry of Economy and Finance |
| MOHW | Ministry of Health and Welfare |
| MOJ | Ministry of Justice |
| MOIS | Ministry of the Interior and Safety |
| MOLIT | Ministry of Land, Infrastructure and Transport |
| MSIT | Ministry of Science and ICT |
| NIA | National Information Society Agency |
| RT- PCR | Real-Time Reverse Transcription Polymerase Chain Reaction |
| R&D | Research and Development |
| SMS | Smart Management System |
| WHO | World Health Organization |